\documentclass[conference, 10pt]{IEEEtran}

\usepackage{bm}
\usepackage[usenames,dvipsnames]{xcolor}
\usepackage{srcltx}
\usepackage{psfrag}
\usepackage{epsfig}
\usepackage{graphicx}
\usepackage{epstopdf}
\usepackage{color}
\usepackage[tbtags,sumlimits,nointlimits,reqno]{amsmath}
\usepackage{amssymb}
\usepackage{color}
\usepackage{float}
\usepackage{subfigure}
\usepackage{gensymb}

\begin{document}

\title{Perfectly Refractive Metasurface using Bianisotropy\vspace{-3mm}}

\author{\IEEEauthorblockN{Guillaume Lavigne, Karim Achouri, and Christophe Caloz}
\IEEEauthorblockA{Department of Electrical Engineering\\
Polytechnique Montr\'{e}al \\
Montr\'{e}al, Qu\'{e}bec H3T 1J4, Canada\vspace{-1cm}}
\and
\IEEEauthorblockN{Viktar Asadchy and Sergei Tretyakov}
\IEEEauthorblockA{Dept. of Radio Science and Engineering\\
Aalto University\\
POB 13000, FI-00076 AALTO, Finland}}

\maketitle


%
\IEEEpeerreviewmaketitle

\begin{abstract}
A passive and reciprocal perfectly refractive metasurface is designed using a general susceptiblity synthesis method. The metasurface uses weak spatial dispersion in the form of bianisotropy to achieve perfect refraction. Its operation is validated using a full-wave simulation of the metasurface implemented with cascaded metallic layers separated by dielectric.
\end{abstract}

\section{Introduction}\label{sec:intro}\vspace{-2mm}

Metasurfaces have recently generated considerable interest because of their ability to control electromagnetic fields with an unprecedented level of flexibility along with their easy fabrication and low-loss characteristics.

One of the most fundamental operations that can be performed with a metasurface is generalized refraction~\cite{yu2011light}. This operation may be essentially achieved using periodic phase-gradient metasurfaces.  However, such rudimentary metasurfaces have been recently shown to suffer from fundamental limitations, related to energy conservation, that preclude their capability to realize ``perfect'' refraction~\cite{asadchy2016perfect,wong2016reflectionless,estakhri2016wave}.

Specifically, Asadchy {\it et al.} have shown in~\cite{asadchy2016perfect} that metasurface perfect refraction is attainable under either of the following three conditions: 1)~the metasurface is nonreciprocal, 2)~the metasurface includes a combination of loss and gain, 3)~the metasurface is bianisotropic with off-diagonal susceptibility tensor. Among these conditions, the third one is probably the most attractive since it corresponds to purely passive and reciprocal designs.

We present here an approach to realize a perfect-refraction bianisotropic metasurface, using the general synthesis technique introduced in~\cite{achouri2014general} and applied in~\cite{achouri2015synthesis}, and demonstrate such a metasurface in a practical design via numerical simulations.

\section{Bianisotropic Metasurface Synthesis}\vspace{-2mm}

The synthesis method in~\cite{achouri2014general} determines the metasurface susceptibility tensors required to transform a specified incident field into specified reflected and transmitted fields using Generalized Sheet Transition Conditions~(GSTCs). In the case of a bianisotropic metasurface lying at $z=0$ in the $xy$-plane of a cartesian coordinate system, the GSTCs are given as
\begin{subequations}\label{eq:1}
\begin{equation}
\tilde{z} \times \Delta\mathbf{H} = j \omega \epsilon \overline{\overline{ \chi}}_\text{ee} \mathbf{E}_\text{av} +  j \omega \overline{\overline{ \chi}}_\text{em} \sqrt{\mu \epsilon}  \mathbf{H}_\text{av} ,
\end{equation}
\begin{equation}
\Delta \mathbf{E} \times \tilde{z}   = j \omega \mu \overline{\overline{ \chi}}_\text{mm} \mathbf{H}_\text{av} +  j \omega \mu \overline{\overline{ \chi}}_\text{me} \sqrt{\epsilon/\mu}  \mathbf{E}_\text{av} ,
\end{equation}
\end{subequations}
where the $\Delta$'s and the `av' subscripts represent the differences and averages, respectively, of the electric or magnetic fields at both sides of the metasurface, and $\overline{\overline{ \chi}}_\text{ee}$, $\overline{\overline{ \chi}}_\text{em}$, $\overline{\overline{ \chi}}_\text{me}$ and $\overline{\overline{ \chi}}_\text{mm}$ are the bianisotropic susceptibility tensors.

We assume that the bianisotropic metasurface is non-gyrotropic, non-chiral, and that its susceptibility tensors are purely transverse. As a result of the former assumption, we may consider purely $x$-polarized fields, since the $x$- and $y$-polarized problems are independent from each other. Under those assumptions and subsequent polarization restriction, the susceptibility tensors, initially including each $3\times 3=9$ components for a total of $9\times 4=36$ components, reduce to tensors with only $\chi^{xx}_\text{ee}$, $\chi^{xy}_\text{em}$, $\chi^{yx}_\text{me}$ and $\chi^{yy}_\text{mm}$ as non-zero components.

\begin{figure}[h]\vspace{-7mm}
\centering
\includegraphics[width=0.75\columnwidth]{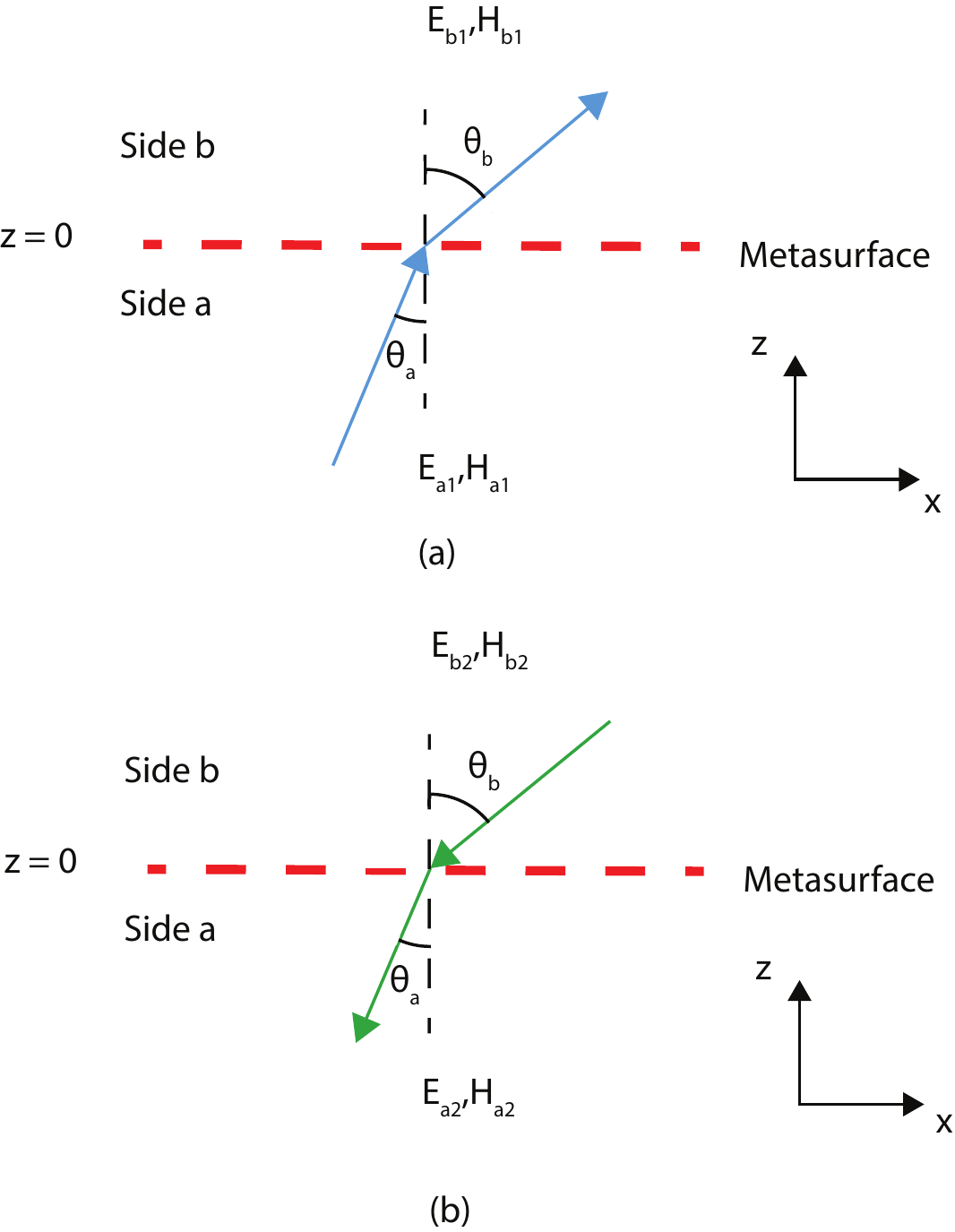}
\caption{The two refraction operations defining the perfect-refraction bianisotropic metasurface. (a)~Direct refraction (transformation~1). (b)~Reciprocal refraction (transformation~2).}\label{fig:transfs}
\end{figure}

Thus, Eqs.~\eqref{eq:1}, once the fields have been specified for a desired refraction, represent a system of 2~equations in 4~unknowns, $\chi^{xx}_\text{ee}$, $\chi^{xy}_\text{em}$, $\chi^{yx}_\text{me}$ and $\chi^{yy}_\text{mm}$. This is an undetermined system, with 2 available degrees of freedom. Now, as mentioned in Sec.~\ref{sec:intro}, we wish, for fabrication simplicity, the metasurface to be reciprocal. This requires the refraction to be symmetric with respect to the direction of propagation. In other words, denoting the two media surrounding the metasurface $a$ and $b$, the refraction from $a$ to $b$ and that from $b$ to $a$ involve exactly the same incidence and transmission pair $(\theta_a,\theta_b)$, as shown in Fig.~\ref{fig:transfs}.

From this point, it is understood that the problem consists in simultaneously synthesizing the two transformations illustrated in Fig.~\ref{fig:transfs}. For the case of TE$_y$ polarization (non-zero field components $E_x$, $E_z$ and $H_y$), the fields may be expressed by the following compact specification, which simultaneously enforces~\eqref{eq:1} for the two transformations:
\begin{equation}\label{eq:two_transf_syst}
\begin{bmatrix}
    \Delta H_{y1}    &      \Delta H_{y2} \\
    \Delta E_{x1}     &      \Delta E_{x2} \\
\end{bmatrix}
=
\begin{bmatrix}
    \chi^{xx}_\text{ee}      & \chi^{xy}_\text{em}  \\
    \chi^{yx}_\text{me}       & \chi^{yy}_\text{mm} \\
\end{bmatrix}
\begin{bmatrix}
          E_{x1,\text{av}}      &   E_{x2,\text{av}} \\
      H_{y1,\text{av}} & H_{y2,\text{av}}
\end{bmatrix}.
\end{equation}

In this system, the first columns of the two field matrices are formed from the fields in the direct transformation [Fig.~\ref{fig:transfs}~(a)], which are (at $z=0$)
\begin{subequations}\label{eq:dir_transf}
\begin{equation}
E_{ax1} = (k_{az}/k_a)e^{-jk_{ax}x} ,\quad H_{ay1} = e^{-jk_{xa}x}/\eta_0,
\end{equation}
\begin{equation}
E_{bx1} = (T k_{bz}/k_b)e^{-jk_{bx}x} ,\quad H_{by1} = (T e^{-jk_{bx}x})/\eta_0,
\end{equation}
\end{subequations}
where $T$ is the transmission coefficient and where $k_{(a,b)x} = k_{a,b} \sin\theta_{a,b}$ and $k_{(a,b)z} = k_{a,b} \cos\theta_{a,b}$
Since we wish a metasurface that is also without gain and loss, we also have to enforce power (or the normal component of the  Poynting vector) conservation, which imposes the following restriction on the transmission coefficient $T=\sqrt{k_{az}/k_{bz}}$.
Note that $T\neq 1$ does \emph{not} mean that the metasurface produces any reflection, dissipation, or gain. This rather expresses the fact that the fields on both sides are different if the incidence and refraction angles are different (in magnitude).

We may now specify the fields for the reciprocal transformation [Fig.~\ref{fig:transfs}(b)]:
\begin{subequations}\label{eq:rec_transf}
\begin{equation}
E_{ax2} = (-k_{az}/k_a) e^{jk_{ax}x}/T ,\quad H_{ay2} = e^{jk_{ax}x}/(T \eta_0),
\end{equation}
\begin{equation}
E_{bx2} = (-k_{bz}/k_b)e^{jk_{bx}x} ,\quad H_{by2} = e^{jk_{bx}x}/\eta_0.
\end{equation}
\end{subequations}
Inserting~\eqref{eq:dir_transf} and~\eqref{eq:rec_transf} into~\eqref{eq:two_transf_syst} finally yields the sought after susceptibility functions $\chi^{xx}_\text{ee}$, $\chi^{xy}_\text{em}$, $\chi^{yx}_\text{me}$ and $\chi^{yy}_\text{mm}$.

\section{Simulation Results}\vspace{-2mm}

Based on the synthesis presented in the previous section, a bianisotropic refractive metasurface is designed to perfectly refract a normally incident wave under the angle of $56^\circ$. The metasurface has been implemented with unit cells composed of three cascaded metallic layers separated by Rogers 3003 substrates. We used dogbone-shaped unit-cell metallic layers, as shown in Fig.~2(a). The metasurface was discretized into 6~different unit cells of size $\lambda_0/5$ and thickness $\lambda_0/10$ at the design frequency of 10~GHz. Each unit cell was then optimized separately in a periodic environment using the commercial software (CST Studio 2014). Similar techniques to realize the scattering particles were also used in~\cite{achouri2015synthesis}.

The six different unit cells are then combined in a supercell which is periodically repeated to form the whole metasurface. This supercell is then be optimized to take into account the different couplings between adjacent unit cells. This is done by simulating the Floquet modes of the whole supercell and modifying the characteristics of the unit cells so as to minimize the power sent into undesired spatial modes. The front-view of the final supercell is shown in Fig.~2(b) while Fig.~2(c) shows the full-wave simulation of the metasurface.
\begin{figure}[h]
\begin{center}
\includegraphics[width=0.75\columnwidth,height=7cm,keepaspectratio]{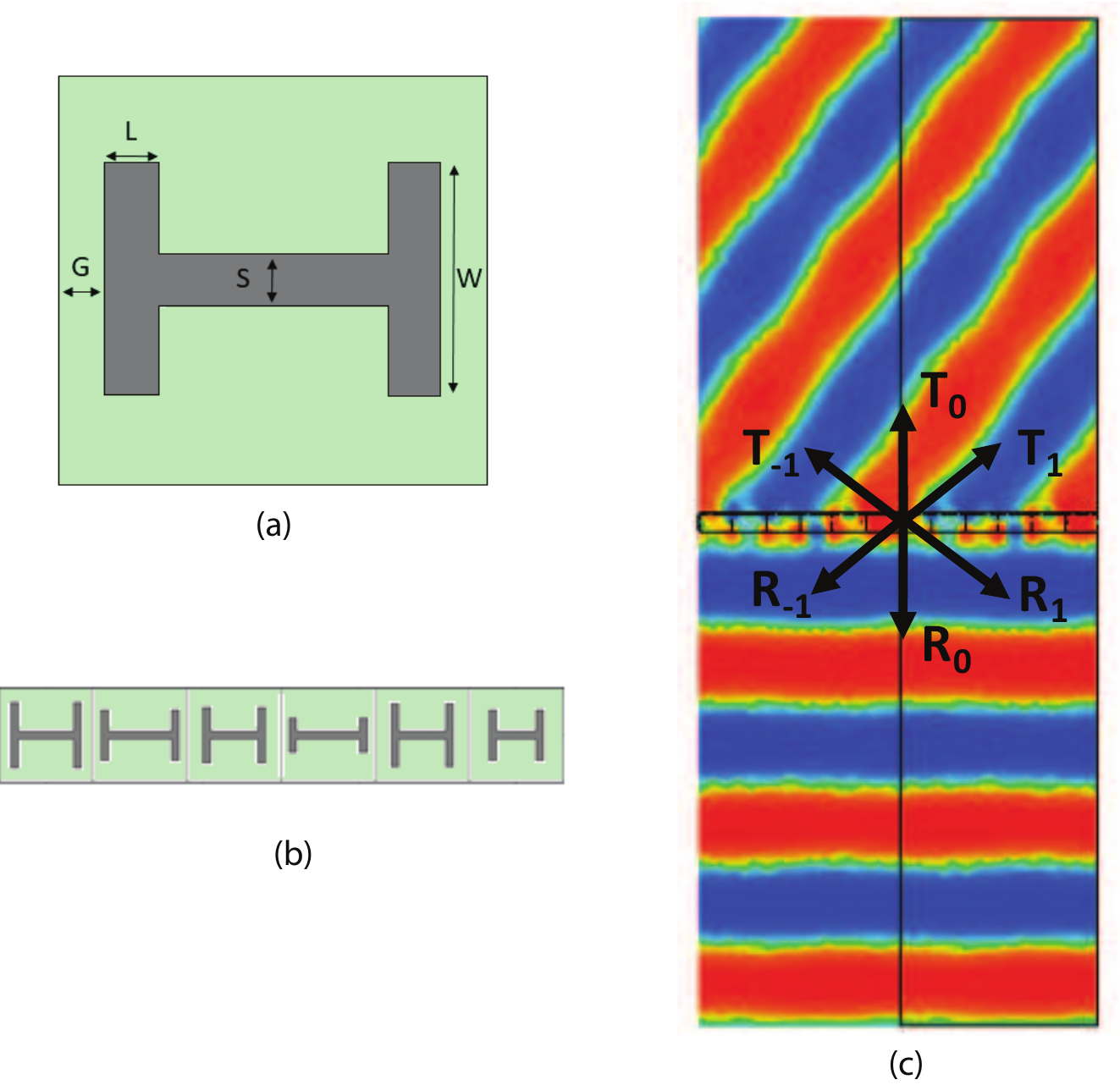}
\caption{Designed metasurface. (a)~Generic unit-cell metallic layer. (b)~Supercell (top layer only visible). (c)~Full-wave simulated fields.}\vspace{-7mm}
\end{center}
\end{figure}

Figure~3 plots the full-wave simulated scattering parameters of the different spatial modes (or diffraction orders). As per specification and optimization, the $T_{-1}$ mode achieves near-perfect transmission ($-0.0953$ dB) near 10.4~GHz with isolation of over 20~dB with respect to the other spatial modes. A similar approach based on three layers of reactive sheets was presented in paper~\cite{epstein2016arbitrary}.

\begin{figure}[h]
\begin{center}
\includegraphics[width=0.75\columnwidth]{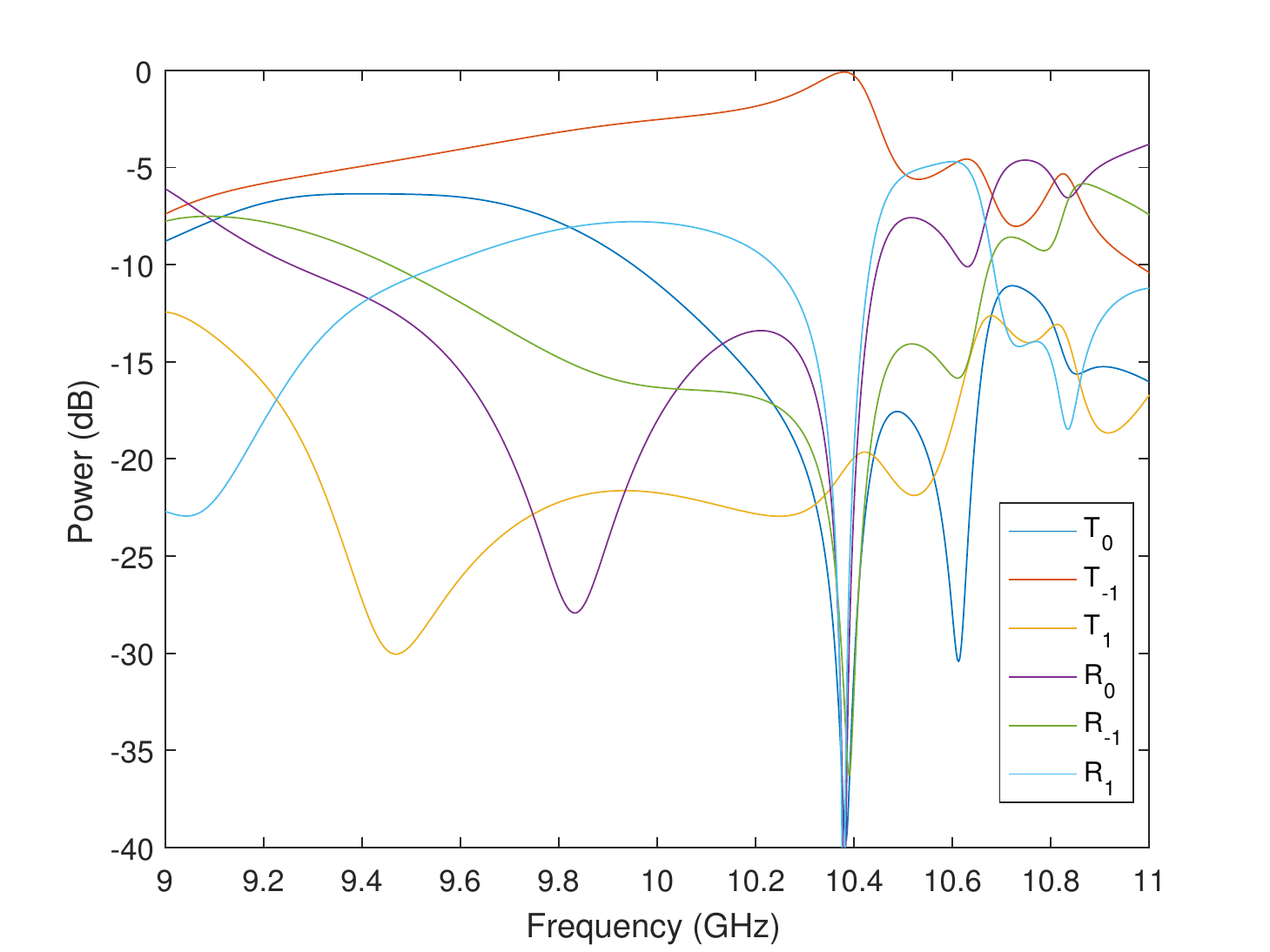}
\caption{Full-wave simulated scattering parameters corresponding to the modes indicated in Fig.~2.}
\end{center}
\end{figure}\vspace{-7mm}

\section{Conclusion}\vspace{-2mm}

We have presented a general technique to realize a passive and reciprocal perfectly refractive metasurface based on a bianisotropic susceptibility tensor synthesis. The metasurface has been demonstrated by full-wave simulation and an experimental design will be presented at the conference.

\small
\bibliographystyle{IEEEtran}
\bibliography{LIB}

\end{document}